\documentclass[amssymb,prb,floats,aps,superscriptaddress,twocolumn,narrowtext]{revtex4}
\usepackage{amsmath}
\usepackage{exscale}
\usepackage[mathscr]{eucal}
\usepackage{epsfig}

\begin{document}

\title{Breakdown of the Wiedemann-Franz law in strongly-coupled
electron-phonon system, application to the cuprates}

\author{K. K. Lee}
\affiliation{IRC in Superconductivity, Cavendish Laboratory,
University of Cambridge, Cambridge, CB3 0HE, United Kingdom}
\author{A. S. Alexandrov}
\affiliation{Department of Physics, Loughborough University,
Loughborough LE11 3TU, United Kingdom}
\author{W. Y. Liang}
\affiliation{IRC in Superconductivity, Cavendish Laboratory,
University of Cambridge, Cambridge, CB3 0HE, United Kingdom}

\begin{abstract}
With the superconducting cuprates in mind, a set of unitary
transformations was used to decouple electrons and phonons in the
strong-coupling limit. While phonons remain almost unrenormalised,
electrons are transformed into itinerent singlet and triplet
bipolarons and thermally excited polarons. The triplet/singlet
exchange energy and the binding energy of the bipolarons are
thought to account for the spin and charge pseudogaps in the
cuprates, respectively. We calculated the Hall Lorenz number of
the system to show that the Wiedemann-Franz law breaks down due to
the interference of the polaron and bipolaron contributions to
heat flow. The model provides a quantitative fit to magnetotransport
data in the cuprates. Furthermore we are able to extract the
phonon component of the thermal conductivity with the use of
experimental data and the model. Our results further validate the
use of a charged Bose gas model to describe normal and
superconducting properties of unconventional superconductors.
\end{abstract}

\vskip2pc

\narrowtext

\bigskip
\maketitle

\section{Introduction}

The theory of high temperature superconductivity is one of the biggest
challenges in condensed matter physics today and is reflected by the large
number of the theoretical models\cite{CHAK,SANF,KIVE,PWA,ZHAN,PINE} proposed
to date. Most of these theories have a foundation built on the
unconventional Cooper pairs whilst others concentrate on the basic
principals of the symmetry breaking and competing orders. On the other hand
the electron-phonon interaction continue to gather support through isotope
effect measurements\cite{ZHAO}, infrared \cite{mic,ita,TIM} and thermal
conductivity\cite{COHN}, neutron scattering \cite{ega}, and more recently in
ARPES\cite{LANZ,CHAI}. To account for the high values of T$_{c}$ in the
cuprates, it is necessary to have electron-phonon interactions larger than
those found in the intermediate coupling theory of superconductivity\cite
{ELIA}. Regardless of the adiabatic ratio, the Migdal-Eliashberg theory of
superconductivity and Fermi-liquids has been shown to breakdown at $\lambda
=1$ \cite{ALEX} using the ($1/\lambda $) expansion technique\cite{LANG}. The
many-electron system collapses into the small (bi)polaron regime\cite
{ALEX,SASH,SAS} at $\lambda \geq 1$ with well separated vibration and
charge-carrier degrees of freedom.

Remarkably the discovery of high T$_{c}$ cuprates was
motivated by the polaron model \cite{MULL}. The isotope effect, high values of
the static dielectric constants and optical spectroscopy in the
cuprates suggest that the electron-phonon interaction is
sufficient to bind small polarons into small bipolarons. At first
sight these carriers have a mass too large to be mobile, however
it has been shown that the inclusion of the on-site Coulomb
repulsion leads to the favoured binding of intersite oxygen
holes\cite {ALEXAND, SANF}. The intersite bipolarons can then
tunnel with an effective mass of only 10 electron
masses\cite{ALEXAND,CATL,alekor,tru}.

Mott and one of the authors (ASA) proposed a simple
model\cite{MOTT} of the cuprates based on bipolarons, Fig.1. In
this model all the holes (polarons) are bound into small intersite
singlet and triplet bipolarons at any temperature. Above T$_{c}$
this Bose gas is non-degenerate and below T$_{c}$ phase coherence
(ODLRO) of the preformed bosons sets in, followed by superfluidity
of the charged carriers. The triplet and singlet states are
separated by an exchange energy J which explains the spin gap
observed in many NMR and neutron scattering
experiments\cite{ROSS,MOOK}. Of course, there are also thermally
excited single polarons in the model. Their density becomes
comparable with the bipolaron density at the temperature $\
T^{\ast } $ which is about half of the bipolaron binding energy
$\Delta $. There is much evidence for the crossover regime at
$T^{\ast }$ and normal state pseudogap in the cuprates
\cite{mickab}. For example, measurements of single particle
tunneling and Andreev reflections\cite{DEUT} clearly demonstrate
that there exists two distinct energy gaps. These energy gaps are
half of the binding energy of preformed bipolarons and the
coherence energy gap of the superconducting phase as proposed in
Ref.\cite{AND}.

Other experimental observations which have been satisfactorily explained
using this particular approach include Hall ratio\cite{BRAT}, Hall angle\cite
{BRAT}, in-plane resistivity\cite{BRAT}, infrared conducitivity\cite{SALJ},
magnetic susceptibility\cite{MULLE}, tunneling spectroscopy\cite{NDROV},
isotope effect\cite{ALEXA}, thermal conducitivity\cite{NEV} and the upper
critical field\cite{SASHA} in the superconducting state. ARPES measurements
indicate the presence of an angular dependent narrow peak and a featureless
background. In the polaronic model the ARPES spectrum can be explained if
one considers a charge transfer Mott-insulator and the single polaron
spectral function\cite{DENT}. The conclusion is that the ARPES spectrum can
in fact be explained with the absence or with a small Fermi surface.
Specific heat measurements have a striking resemblance to the superfluid
transition of weakly-interacting 3D bosons\cite{JUN}. A theoretical model
\cite{DROV}, experimental data and analysis of the specific heat in a
magnetic field\cite{JUNO} suggest this is a convincing argument.

\begin{figure}[h]
\centering \epsfig{file=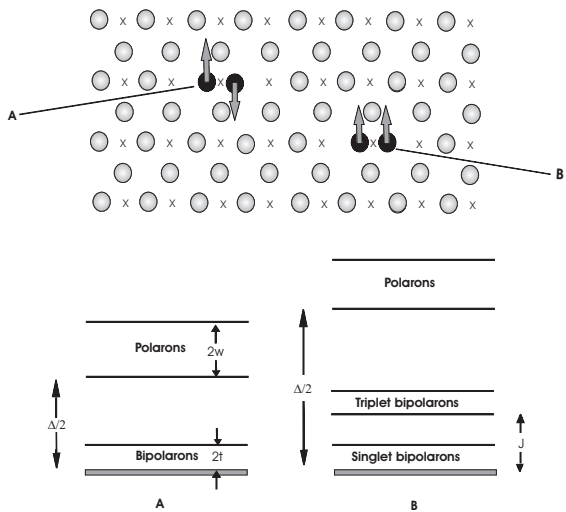, bbllx=50, bblly=308, bburx=561,
bbury=702,width=3.5in} \caption{Bipolaron picture of high
temperature superconductors. $A$ corresponds to the singlet
intersite bipolaron. $B$ is the triplet intersite bipolaron, which
naturally includes the addition of an extra excitation band. The
crosses are copper sites and the circles are oxygen sites. w is
the half bandwidth of the polaron band, t is the half bandwidth of
the bipolaron band, $\Delta/2$ is the bipolaron binding energy per
polaron and J is the exchange energy per bipolaron.} \label{fig.1}
\end{figure}

In this paper we develop the theory of the normal state magnetotransport in
bipolaronic systems and explain the experimental Hall Lorenz number for $%
YBa_{2}Cu_{3}O_{6+x}$ as measured by Zhang et al.\cite{ZHANG}. The case of
degenerate singlet and triplet bipolarons was briefly discussed by us in
Ref. \cite{kkl}. Particular interest within this paper lies in the
conclusion that the Wiedemann-Franz law is violated in cuprates. This
departure from the Fermi liquid picture is seen in both the superconducting
and normal state and might be related by a common mechanism\cite{TAKE}.

The extraction of the electronic thermal conductivity term has proven
difficult as both the electronic term, $\kappa _{el}$ and the phononic term,
$\kappa _{ph}$ make a comparable contribution. However recent experiments
appear to have got around this problem\cite{SALA,TAKE,ZHANG}. Takenaka et
al. \cite{TAKE} found that the insulating state thermal conductivity
(therefore predominantly phononic) was approximately independent of doping
except at y $\sim 1-0.75$ in $YBa_{2}Cu_{3}O_{7-y}$. Therefore the phononic
term for the thermal conductivity above $y\sim 0.5$ could be estimated. This
analysis led to the conclusion that the $\kappa _{el}$ is constant or weakly
T-dependent in the normal state like that found by Salamon et al\cite{SALA}.
This approximately T-independent $\kappa _{el}$ in the underdoped region
therefore implies the violation of the Wiedemann-Franz law (since the
resistivity is found to be a non-linear function of temperature in this
regime). This breakdown of the Wiedemann-Franz law has been seen in other
cuprates such as $Pr_{2-x}Ce_{x}CuO_{4}$ at optimal doping\cite{HILL} and $%
La_{2-x}Sr_{x}CuO_{4}$ at underdoping \cite{ANDO}. Zhang et al\cite{ZHANG}
recently developed a novel method to determine the Lorenz number, based on
the thermal Hall conductivity. The thermal Hall effect is a purely
electronic effect since phonons are not affected by the magnetic field. As a
result, the ``Hall'' Lorenz number, $L_{H}=\left( e/k_{B}\right) ^{2}\kappa
_{xy}/(T\sigma _{xy})$, has been directly measured in $YBa_{2}Cu_{3}O_{6.95}$
because transverse thermal $\kappa _{xy}$ and electrical $\sigma _{xy}$
conductivities involve only the electrons. The experimental $L_{xy}$ showed
a linear temperature dependence, which violates the Wiedemann-Franz law. It
is clear that it would be difficult to explain this experimental observation
in the framework of any Fermi-liquid model.

A charge 2e Bose gas of bipolarons naturally leads to the conclusion\cite
{NEV} that the Lorenz number in the normal state is strongly suppressed due
to the double charge per carrier and classical statistics of bipolarons
compared with the Lorenz number in the normal metal,
$L_{0}=\pi^{2}/3$.
Here we demonstrate that the Wiedemann-Franz law breaks
down because of the interference of the polaron and bipolaron contributions
to the heat transport. When thermally excited polarons and triplet pairs are
included, the bipolaron model explains the violation of the Wiedemann-Franz
law in the cuprates and the Hall Lorenz number as seen in the experiments.

\section{Low Fermi energy: pairing is individual in many cuprates}

The possibility of real-space pairing, as opposed to BCS-like pairing, has
been the subject of much discussion. Experimental and theoretical evidence
for an exceptionally strong electron-phonon interaction in high temperature
superconductors is now so overwhelming, that even advocates of nonphononic
mechanisms \cite{kiv} accept this fact. Nevertheless the same authors \cite
{KIVE,kiv} dismiss any real-space pairing claiming that pairing is
collective in cuprates. They believe in a large Fermi surface with the
number of holes $(1+x)$ rather than $x$ in superconducting cuprates, where $%
x $ is the doping level like in $La_{2-x}Sr_{x}CuO_{4}.$ As an alternative
to a three-dimensional Bose-Einstein condensation of bipolarons these
authors suggest a collective pairing (i.e the{\it \ Cooper }pairs in the
momentum space) at some temperature $T^{\ast }>T_{c},$ but without phase
ordering. In this concept the phase coherence and superconducting critical
temperature $T_{c}$ are determined by the superfluid density, which is
proportional to doping $x$ due to a low dimensionality, rather then to the
total density of carriers ($1+x)$. On the experimental side a large Fermi
surface is clearly incompatible with a great number of thermodynamic,
magnetic, and kinetic measurements, which show that only holes {\it doped }%
into a parent insulator are carriers in the {\it normal} state. On the
theoretical ground this preformed {\it Cooper}-pair (or phase-fluctuation)
scenario contradicts a theorem \cite{pop}, which proves that the number of
supercarriers (at $T=0$) and normal-state carriers is the same in any {\it %
clean} superfluid.

Objections against real-space pairing also contradict a parameter-free
estimate of the Fermi energy, which shows that the pairing is individual in
cuprates \cite{aleF}. Indeed, first-principles band structure calculations
show that copper, alkali metal, and magnesium donate their outer electrons
to oxygen, $C_{60},$ and boron in cuprates, doped fullerenes, and in $%
MgB_{2} $, respectively. In cuprates the band structure is
quasi-two-dimensional (2D) with a few degenerate hole pockets. Applying the
parabolic approximation for the band dispersion we obtain the {\it %
renormalized} Fermi energy as
\begin{equation}
\epsilon _{F}={\frac{\hbar ^{2}\pi n_{i}d}{{m_{i}^{\ast }}}},
\end{equation}
where $d$ is the interplane distance, and $n_{i},m_{i}^{\ast }$ are the
density of holes and their effective mass in each of the hole subbands $i$
renormalized by the electron-phonon (and by any other) interaction. One can
express the renormalized band-structure parameters through the in-plane
magnetic-field penetration depth at $T\approx 0$, measured experimentally:
\begin{equation}
\lambda _{H}^{-2}=4\pi e^{2}\sum_{i}\frac{n_{i}}{m_{i}^{\ast }c^{2}}.
\end{equation}
As a result, we obtain a {\em parameter-free }expression for the ``true''
Fermi energy as
\begin{equation}
\epsilon _{F}={\frac{\hbar ^{2}dc^{2}}{{4ge^{2}\lambda _{H}^{2}}}},
\end{equation}
where $g$ is the degeneracy of the spectrum which might depend on doping in
cuprates. One expects $4$ hole pockets inside the Brillouin zone (BZ) due to
the Mott-Hubbard gap in underdoped cuprates. If the hole band minima are
shifted with doping to BZ boundaries, all their wave vectors would belong to
the stars with two or more prongs. The groups of wave vectors for these
stars have only 1D representations. It means that the spectrum will be
degenerate with respect to the number of prongs which the star has, i.e $%
g\geqslant 2$. Because Eq.(3) does not contain any other band-structure
parameters, the estimate of $\epsilon _{F}$ using this equation does not
depend very much on the parabolic approximation for the band dispersion.

Generally, the ratios $n/m^{\ast }$ in Eq.(1) and Eq.(2) are not
necessarily the same. The `superfluid' density in Eq.(2) might be
different from the total density of delocalized carriers in
Eq.(1). However, in a translationally invariant system they must
be the same \cite{pop}. This is also true even in the extreme case
of a pure two-dimensional superfluid, where quantum fluctuations
are important. One can, however, obtain a reduced value of the
zero temperature superfluid density in the dirty limit, $l\ll \xi
(0)$, where $\xi (0)$ is the zero-temperature coherence length.
The latter was measured directly in cuprates as the size of the
vortex core. It is about 10 $\AA $ or even less. On the contrary,
the mean free path was found surprisingly large at low
temperatures, $l\sim $ 100-1000 $\AA $. Hence, it is rather
probable that novel superconductors are in the clean limit, $l\gg
\xi (0)$, so that the parameter-free expression for $\epsilon
_{F}$, Eq.(3) holds.

Parameter-free estimates of the Fermi energy obtained by using Eq.(3) are
presented in the Table.
\begin{table}[tbp]
\caption{The Fermi energy (multiplied by the degeneracy) of cuprates}
\begin{tabular}[t]{llllllll}
Compound & $T_{c}$ (K) & $\lambda _{H,ab}$ $(\AA )$ & d$(\AA )$ & $%
g\epsilon_{F}$ (meV) &  &  &  \\ \hline
$La_{1.8}Sr_{0.2}CuO_{4}$ & 36.2 & 2000 & 6.6 & 112 &  &  &  \\
$La_{1.78}Sr_{0.22}CuO_{4}$ & 27.5 & 1980 & 6.6 & 114 &  &  &  \\
$La_{1.76}Sr_{0.24}CuO_{4}$ & 20.0 & 2050 & 6.6 & 106 &  &  &  \\
$La_{1.85}Sr_{0.15}CuO_{4}$ & 37.0 & 2400 & 6.6 & 77 &  &  &  \\
$La_{1.9}Sr_{0.1}CuO_{4}$ & 30.0 & 3200 & 6.6 & 44 &  &  &  \\
$La_{1.75}Sr_{0.25}CuO_{4}$ & 24.0 & 2800 & 6.6 & 57 &  &  &  \\
$YBa_{2}Cu_{3}O_{7}$ & 92.5 & 1400 & 4.29 & 148 &  &  &  \\
$YBaCuO(2\%Zn)$ & 68.2 & 2600 & 4.29 & 43 &  &  &  \\
$YBaCuO(3\%Zn)$ & 55.0 & 3000 & 4.29 & 32 &  &  &  \\
$YBaCuO(5\%Zn)$ & 46.4 & 3700 & 4.29 & 21 &  &  &  \\
$YBa_{2}Cu_{3}O_{6.7}$ & 66.0 & 2100 & 4.29 & 66 &  &  &  \\
$YBa_{2}Cu_{3}O_{6.57}$ & 56.0 & 2900 & 4.29 & 34 &  &  &  \\
$YBa_{2}Cu_{3}O_{6.92}$ & 91.5 & 1861 & 4.29 & 84 &  &  &  \\
$YBa_{2}Cu_{3}O_{6.88}$ & 87.9 & 1864 & 4.29 & 84 &  &  &  \\
$YBa_{2}Cu_{3}O_{6.84}$ & 83.7 & 1771 & 4.29 & 92 &  &  &  \\
$YBa_{2}Cu_{3}O_{6.79}$ & 73.4 & 2156 & 4.29 & 62 &  &  &  \\
$YBa_{2}Cu_{3}O_{6.77}$ & 67.9 & 2150 & 4.29 & 63 &  &  &  \\
$YBa_{2}Cu_{3}O_{6.74}$ & 63.8 & 2022 & 4.29 & 71 &  &  &  \\
$YBa_{2}Cu_{3}O_{6.7}$ & 60.0 & 2096 & 4.29 & 66 &  &  &  \\
$YBa_{2}Cu_{3}O_{6.65}$ & 58.0 & 2035 & 4.29 & 70 &  &  &  \\
$YBa_{2}Cu_{3}O_{6.6}$ & 56.0 & 2285 & 4.29 & 56 &  &  &  \\
$HgBa_{2}CuO_{4.049}$ & 70.0 & 2160 & 9.5 & 138 &  &  &  \\
$HgBa_{2}CuO_{4.055}$ & 78.2 & 1610 & 9.5 & 248 &  &  &  \\
$HgBa_{2}CuO_{4.055}$ & 78.5 & 2000 & 9.5 & 161 &  &  &  \\
$HgBa_{2}CuO_{4.066}$ & 88.5 & 1530 & 9.5 & 274 &  &  &  \\
$HgBa_{2}CuO_{4.096}$ & 95.6 & 1450 & 9.5 & 305 &  &  &  \\
$HgBa_{2}CuO_{4.097}$ & 95.3 & 1650 & 9.5 & 236 &  &  &  \\
$HgBa_{2}CuO_{4.1}$ & 94.1 & 1580 & 9.5 & 257 &  &  &  \\
$HgBa_{2}CuO_{4.101}$ & 93.4 & 1560 & 9.5 & 264 &  &  &  \\
$HgBa_{2}CuO_{4.101}$ & 92.5 & 1390 & 9.5 & 332 &  &  &  \\
$HgBa_{2}CuO_{4.105}$ & 90.9 & 1560 & 9.5 & 264 &  &  &  \\
$HgBa_{2}CuO_{4.108}$ & 89.1 & 1770 & 9.5 & 205 &  &  &  \\ \hline
\end{tabular}
\end{table}
The renormalised Fermi energy in more than 30 cuprates is definitely less
than $100$ $meV$.

In many cases (Table) the renormalized Fermi energy is so small that pairing
is certainly individual. Such pairing will occur when the size of a pair, $%
\rho $ is smaller than the inter-pair separation, $r$. The size of a pair is
generally
\begin{equation}
\rho =\frac{\hbar }{\sqrt{m^{\ast }\Delta }},
\end{equation}
where $\Delta $ is its binding energy. The separation of pairs can be
directly related to the Fermi energy in 2D
\begin{equation}
r=\hbar \sqrt{\frac{\pi }{\epsilon _{F}m^{\ast }}}.
\end{equation}
We see that the true condition for real-space pairing is
\begin{equation}
\epsilon _{F}\lesssim \pi \Delta.
\end{equation}

The bipolaron binding energy is thought to be twice the so-called
pseudogap \cite{SANF}. Experimentally measured pseudogap of many
cuprates \cite{mickab} is as large as $\Delta /2\gtrsim 50meV,$ so
that Eq.(6) is well satisfied in underdoped and even in a few
optimally and overdoped cuprates. One should notice that the
coherence length in the charged Bose gas has nothing to do with
the size of the boson as erroneously assumed by some authors
\cite{kiv}. It depends on the interparticle distance and the
mean-free path, and might be as large as in the BCS
superconductor. Hence, it would be incorrect to apply the ratio of
the coherence length to the inter-carrier distance as a criterium
of the BCS-Bose liquid crossover. The correct criterium is given
by Eq.(6).

\section{Non-equilibrium bipolaron and polaron distribution functions}

Because thermally excited phonons and (bi)polarons are well decoupled in the
strong-coupling regime of the electron-phonon interaction the standard
Boltzmann equation for kinetics of renormalised carriers may be applied. The
distribution function for each carrier is given as
\begin{equation}
f=f_{0}+f_{1}.  \label{1}
\end{equation}
Here $f_{0}$ is the distribution function at equilibrium and $f_{1}({\bf k})$
is the deviation of the distribution function away from equilibrium. We
assume that $f_{1}$ is small compared to $f_{0}$. $f_{0}^{p}$ (polaron) is
the Fermi-Dirac distribution and both $f_{0}^{s}$ and $f_{0}^{t}$ (singlet
and triplet bipolaron, respectively) are the Bose-Einstein distributions,
\begin{equation}
f_{0}^{p}={\frac{1}{{\exp [(}E{+\Delta /2-\mu /2)/T]+1}}},  \label{6}
\end{equation}
\begin{equation}
f_{0}^{s}={\frac{1}{{\exp [(E-\mu )/T]-1}}},  \label{7}
\end{equation}
\begin{equation}
f_{0}^{t}={\frac{1}{{\exp [(E+J-\mu )/T]-1}}}.  \label{8}
\end{equation}
$\mu $ is the chemical potential, $J$ is the exchange energy which separates
the triplet state from the singlet state, and $\Delta $ is the bipolaron
binding energy per pair which is assumed to be of s-symmetry. The latter
assumption of s-wave bulk pairing symmetry of a single-particle gap has been
shown to be a valid one \cite{mul2,AND,ZHAO1}.

We make use of the $\tau -$approximation\cite{ANSE} in the presence of the
electric field {\bf E}, temperature gradient ${\bf \nabla }{T}$ and magnetic
field {\bf B}$\parallel $ {\bf z} $\perp $ {\bf E} and ${\bf \nabla }{T}$,
\begin{equation}
f({\bf k})=f_{0}(E)+\tau \frac{\partial f_{0}}{\partial E}{\bf v}\cdot \frac{%
\left\{ {\bf F}+g\tau {\bf B}\times {\bf F}\right\} }{1+(g\tau B)^{2}}
\label{2}
\end{equation}
where we have set $\hbar =1$ and also from now on set $k_{B}=c=1$. ${\bf v}%
=\partial E/\partial {\bf k}$, $\tau $ is the relaxation time and we assume
that it depends on the kinetic energy, $E=k^{2}/2m$. ${\bf F}=(E-\mu ){\bf \nabla }T/T+{\bf %
\nabla }(\mu -2e\phi )$ and $g=g_{s}=2e/m_{s}$ for singlet bipolarons with
the energy $E=k^{2}/(2m_{s})$. For triplet bipolarons, ${\bf F}=(E+J-\mu )%
{\bf \nabla }T/T+{\bf \nabla }(\mu -2e\phi )$, $g=g_{t}=2e/m_{t}$ and the
energy $E=k^{2}/(2m_{t})$. ${\bf F}=(E+\Delta /2-\mu /2){\bf \nabla }T/T+%
{\bf \nabla }(\mu /2-e\phi )$, $E=k^{2}/(2m_{p})$ and $g=g_{p}=e/m_{p}$ for
thermally excited polarons. Here $m_{s,t,p}$ are the singlet and triplet
bipolaron and polaron masses of $two$-dimensional carriers.

Eqns.(7-11) are used to find the thermal and electrical currents induced by
the applied thermal and potential gradients in a magnetic field.

\section{Electrical and thermal currents}

The electrical current for each carrier is given by
\begin{equation}
{\bf j}^{\alpha }{\bf =}q\sum_{{\bf k}}{\bf v}f_{1}^{\alpha }({\bf k})
\end{equation}
where $\alpha =s,p,t$, and $q$ is the carrier charge. We find the x
direction component as
\begin{eqnarray}
j_{x}^{\alpha } &=&a_{xx}^{\alpha }\nabla _{x}(\mu -2e\phi )+a_{xy}^{\alpha
}\nabla _{y}(\mu -2e\phi )  \nonumber \\
&&+b_{xx}^{\alpha }\nabla _{x}T+b_{xy}^{\alpha }\nabla _{y}T
\end{eqnarray}
and also the y direction component
\begin{eqnarray}
j_{y}^{\alpha } &=&a_{yy}^{\alpha }\nabla _{y}(\mu -2e\phi )+a_{yx}^{\alpha
}\nabla _{x}(\mu -2e\phi )  \nonumber \\
&&+b_{yy}^{\alpha }\nabla _{y}T+b_{yx}^{\alpha }\nabla _{x}T
\end{eqnarray}
where
\begin{eqnarray}
a_{xx}^{p} &=&a_{yy}^{p}=\frac{en_{p}}{2m_{p}}\langle \tau _{p}\rangle , \\
a_{yx}^{p} &=&-a_{xy}^{p}=\frac{eg_{p}Bn_{p}}{2m_{p}}\langle \tau
_{p}^{2}\rangle ,  \nonumber \\
b_{xx}^{p} &=&b_{yy}^{p}=\frac{en_{p}}{Tm_{p}}\langle \tau _{p}\{E+\Delta
/2-\mu /2\}\rangle ,  \nonumber \\
b_{yx}^{p} &=&-b_{xy}^{p}=\frac{eg_{p}Bn_{p}}{Tm_{p}}\langle \tau
_{p}^{2}\{E+\Delta /2-\mu /2\}\rangle ,  \nonumber \\
a_{xx}^{s,t} &=&a_{yy}^{s,t}=\frac{2en_{s,t}}{m_{s,t}}\langle \tau
_{s,t}\rangle
,  \nonumber \\
a_{yx}^{s,t}
&=&-a_{xy}^{s,t}=\frac{2eg_{s,t}Bn_{s,t}}{m_{s,t}}\langle \tau
_{s,t}^{2}\rangle ,  \nonumber
\\
b_{xx}^{s} &=&b_{yy}^{s}=\frac{2en_{s}}{Tm_{s}}\langle \tau _{s}\{E-\mu
\}\rangle ,  \nonumber \\
b_{yx}^{s} &=&-b_{xy}^{s}=\frac{2eg_{s}Bn_{s}}{Tm_{s}}\langle \tau
_{s}^{2}\{E-\mu \}\rangle  \nonumber \\
b_{xx}^{t} &=&b_{yy}^{t}=\frac{2en_{t}}{Tm_{t}}\langle \tau _{t}\{E+J-\mu
\}\rangle ,  \nonumber \\
b_{yx}^{t} &=&-b_{xy}^{t}=\frac{2eg_{t}Bn_{t}}{Tm_{t}}\langle \tau
_{t}^{2}\{E+J-\mu \}\rangle ,  \nonumber
\end{eqnarray}
and
\begin{equation}
\langle \tau _{\alpha }^{r}\rangle =\frac{\int_{0}^{\infty }dEE\tau _{\alpha
}^{r}(E)[1+(g_{\alpha }\tau _{\alpha }(E)B)^{2}]^{-1}\partial f_{0}^{\alpha
}/\partial E}{\int_{0}^{\infty }dEf_{0}^{\alpha }}
\end{equation}

The number densities, $n_{\alpha }$ of the three carriers can be evaluated
as
\begin{equation}
n_{p}=\frac{m_{p}T}{\pi }\ln \left[ 1+\exp \left( -\frac{\Delta -\mu }{2T}%
\right) \right] ,  \label{29}
\end{equation}
\begin{equation}
n_{s}=-\frac{m_{s}T}{2\pi }\ln \left[ 1-\exp \left( \frac{\mu }{T}\right) %
\right] ,  \label{30}
\end{equation}
\begin{equation}
n_{t}=-\frac{3m_{t}T}{2\pi }\ln \left[ 1-\exp \left( \frac{\mu -J}{T}\right) %
\right] .  \label{31}
\end{equation}
and the results are shown in Fig. 2 for optimally doped $%
YBa_{2}Cu_{3}O_{6.95} $ using experimental estimates of the energy gaps as
measured by Mihailovic et al \cite{mickab}.

\begin{figure}[h]
\centering \epsfig{file=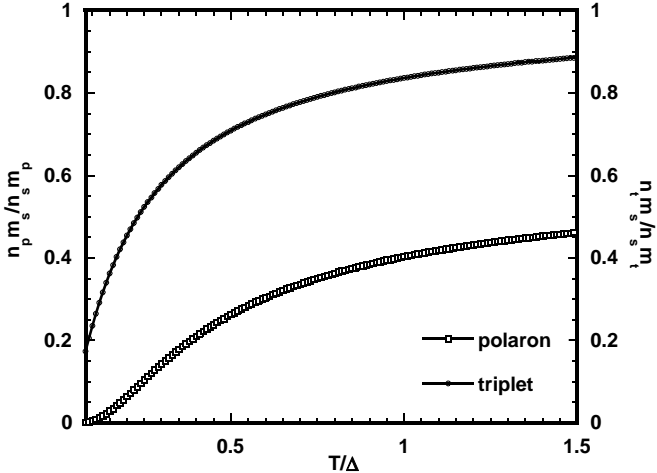,bbllx=78, bblly=223, bburx=543,
bbury=565,width=3.5in} \caption{number density ratios for triplets
and polarons in the optimally doped regime} \label{fig.2}
\end{figure}

To find the thermal currents we calculate the energy flux for each carrier
which are given by
\begin{equation}
{\bf w}^{\alpha }{\bf =}\sum_{{\bf k}}{\bf v}f_{1}^{\alpha }({\bf k}%
)E^{\alpha }  \label{9}
\end{equation}
where $E^{p}=E+\Delta /2+e\phi $, $E^{s}=E+2e\phi $ and $E^{t}=E+J+2e\phi $.
The x direction component is
\begin{eqnarray}
w_{x}^{\alpha } &=&c_{xx}^{\alpha }\nabla _{x}(\mu -2e\phi )+c_{xy}^{\alpha
}\nabla _{y}(\mu -2e\phi )  \nonumber \\
&&+d_{xx}^{\alpha }\nabla _{x}T+d_{xy}^{\alpha }\nabla _{y}T  \label{10}
\end{eqnarray}
and the y direction component is
\begin{eqnarray}
w_{y}^{\alpha } &=&c_{yy}^{\alpha }\nabla _{y}(\mu -2e\phi )+c_{yx}^{\alpha
}\nabla _{x}(\mu -2e\phi )  \nonumber \\
&&+d_{yy}^{\alpha }\nabla _{y}T+d_{yx}^{\alpha }\nabla _{x}T  \label{10}
\end{eqnarray}
where
\begin{eqnarray}
c_{xx}^{p} &=&c_{yy}^{p}=\frac{n_{p}}{2m_{p}}\langle E^p \tau _{p}\rangle , \\
c_{yx}^{p} &=&-c_{xy}^{p}=\frac{g_{p}Bn_{p}}{2m_{p}}\langle E^p \tau
_{p}^{2}\rangle ,  \nonumber \\
d_{xx}^{p} &=&d_{yy}^{p}=\frac{n_{p}}{Tm_{p}}\langle E^p (E+\Delta
/2-\mu /2) \tau _{p}\rangle ,  \nonumber \\
d_{yx}^{p} &=&-d_{xy}^{p}=\frac{g_{p}Bn_{p}}{Tm_{p}}\langle E^p (E+\Delta /2-\mu /2)\tau
_{p}^{2}\rangle ,  \nonumber \\
c_{xx}^{s} &=&c_{yy}^{s}=\frac{n_{s}}{m_{s}}\langle E^s \tau _{s}\rangle ,  \nonumber \\
c_{yx}^{s} &=&-c_{xy}^{s}=\frac{g_{s}Bn_{s}}{m_{s}}\langle E^s \tau
_{s}^{2}\rangle ,  \nonumber \\
d_{xx}^{s} &=&d_{yy}^{s}=\frac{n_{s}}{Tm_{s}}\langle E^s (E-\mu) \ \tau
_{s}\rangle ,  \nonumber \\
d_{yx}^{s} &=&-d_{xy}^{s}=\frac{g_{s}Bn_{s}}{Tm_{s}}\langle E^s
(E-\mu) \tau^{2}_s\rangle ,  \nonumber \\
c_{xx}^{t} &=&c_{yy}^{t}=\frac{n_{t}}{m_{t}}\langle E^t \tau _{t}\rangle ,  \nonumber \\
c_{yx}^{t} &=&-c_{xy}^{t}=\frac{g_{t}Bn_{t}}{m_{t}}\langle E^t \tau
_{t}^{2}\rangle ,  \nonumber \\
d_{xx}^{t} &=&d_{yy}^{t}=\frac{n_{t}}{Tm_{t}}\langle E^t (E+J-\mu) \tau
_{t}\rangle ,  \nonumber \\
d_{yx}^{t} &=&-d_{yx}^{t}=\frac{g_{t}Bn_{t}}{Tm_{t}}\langle E^t
(E+J-\mu) \tau
_{t}^{2}\rangle.  \nonumber
\end{eqnarray}

\section{Transport coefficients}

In thermal transport experiments one has to make sure the
electrical current is set to zero ${\bf j}={\bf j_{p}}+{\bf
j_{s}}+{\bf j_{t}}=0$. Using eqns. (13) and (14), we find,
\begin{eqnarray}
\nabla_x(\mu-2e\phi)=-\frac{b_{yx}a_{yx}+b_{xx}a_{xx}}{(a_{xx})^2+(a_{yx})^2}\nabla_xT
\nonumber\\
+\frac{b_{yx}a_{xx}-b_{xx}a_{yx}}{(a_{xx})^2+(a_{yx})^2}\nabla_yT
\end{eqnarray}
and
\begin{eqnarray}
\nabla_y(\mu-2e\phi)=-\frac{b_{yy}a_{yy}+b_{yx}a_{yx}}{(a_{xx})^2+(a_{yx})^2}\nabla_yT
\nonumber\\
+\frac{b_{xx}a_{yx}-b_{yx}a_{xx}}{(a_{xx})^2+(a_{yx})^2}\nabla_xT
\end{eqnarray}
where $b_{xx}=b_{xx}^p+b_{xx}^s+b_{xx}^t$,
$a_{yx}=a_{yx}^p+a_{yx}^s+a_{yx}^t$, etc. By substituting eqn (24)
and (25) into eqn (21) and (22) we obtain for the thermal
conductivities $\kappa _{xx},$ and for the thermal Hall
conductivity, $\kappa _{yx}$
\begin{eqnarray}
\kappa _{xx} &=&d_{xx}-c_{xx}\frac{b_{yx}a_{yx}+b_{xx}a_{xx}}{(a_{xx})^{2}+(a_{yx})^{2}}  \nonumber \\
&&+c_{yx}\frac{b_{yx}a_{xx}-b_{xx}a_{yx}}{(a_{xx})^{2}+(a_{yx})^{2}}
\end{eqnarray}
and
\begin{eqnarray}
\kappa _{yx} &=&d_{yx}-c_{yx}\frac{b_{yx}a_{yx}+b_{xx}a_{xx}}{(a_{xx})^{2}+(a_{yx})^{2}}  \nonumber \\
&&+c_{xx}\frac{b_{xx}a_{yx}-b_{yx}a_{xx}}{(a_{xx})^{2}+(a_{yx})^{2}}
\end{eqnarray}

The electrical conductivity is defined in the presence of an electric field $%
{\bf E}=-{\bf \nabla }\phi $ and in the absence of a thermal and chemical
gradient, ${\bf \nabla }T={\bf \nabla }\mu =0,$
\begin{equation}
j_{x}=2ea_{xx}E_{x}-2ea_{yx}E_{y}
\end{equation}
and
\begin{equation}
j_{y}=2ea_{yy}E_{y}+2ea_{yx}E_{x}
\end{equation}
which gives
\begin{eqnarray}
\sigma _{xx} &=&\sigma _{yy}=2ea_{xx} \\
\sigma _{yx} &=&-\sigma _{xy}=2ea_{yx}
\end{eqnarray}
We can combine eqns. (26),(27),(30) and (31) to obtain the Lorenz number and
the Hall Lorenz number as
\begin{eqnarray}
L&=&\frac{e}{2Ta_{xx}}(d_{xx}-c_{xx}\frac{b_{yx}a_{yx}+b_{xx}a_{xx}}{(a_{xx})^{2}+(a_{yx})^{2}}\nonumber
\\
&&+c_{yx}\frac{b_{yx}a_{xx}-b_{xx}a_{yx}}{(a_{xx})^{2}+(a_{yx})^{2}}) \\
L_{H}&=&\frac{e}{2Ta_{yx}}(d_{yx}-c_{yx}\frac{b_{yx}a_{yx}+b_{xx}a_{xx}}{(a_{xx})^{2}+(a_{yx})^{2}}  \nonumber \\
&&+c_{xx}\frac{b_{xx}a_{yx}-b_{yx}a_{xx}}{(a_{xx})^{2}+(a_{yx})^{2}}),
\end{eqnarray}
respectively.

We also define the Hall ratio. The Hall effect is the appearance
of an electric field perpendicular to both the current and applied
magnetic field, which is perpendicular to the current. If the
current flows in the x-direction only, $j_{x}=j$ and $j_{y}=0$,
then we can eliminate $E_{x}$ and find the Hall ratio, which is
defined as $R_{H}=E_{y}/(Bj)$,
\begin{equation}
R_{H}=\frac{a_{yx}}{2e[(a_{xx})^{2}+(a_{yx})^{2}]B}
\end{equation}

\section{Weak-field approximation}

If the magnetic field is weak, $g_{\alpha }\tau _{\alpha }B<<1,$ we can
ignore all terms in $B^{2}$ and higher order. The previous definition of $%
\langle \tau _{\alpha }^{r}\rangle $ simplifies and becomes
\begin{equation}
\langle \tau _{\alpha }^{r}\rangle =\frac{\int_{0}^{\infty }E\tau _{\alpha
}^{r}dE\partial f_{0}^{\alpha }/\partial E}{\int_{0}^{\infty
}dEf_{0}^{\alpha }}
\end{equation}
Next we can apply the coefficients, eqns (15) and (23), into the transport
coefficients derived in the previous section.

The resistivity, Hall resistivity and Hall ratio for a triplet,
singlet bipolaron and polaron system are respectively
\begin{eqnarray}
\rho _{xx} &=&\rho _{yy}=\frac{m_p}{e^{2}n_{p}\langle
\tau_{p}\rangle [1+4A_{s1}+4A_{t1}]}
\\
\rho _{yx} &=&-\rho _{xy}=\frac{m_p}{e^{2}g_{p}Bn_{p}\langle
\tau_{p}^{2}\rangle[1+4A_{s2}+4A_{t2}]}
\\
R_{H}&=&\frac{g_p\langle \tau _p^2\rangle
m_p[1+4A_{s2}+4A_{t2}]}{e^2 \langle \tau
_p\rangle^2n_p[1+4A_{s1}+4A_{t1}]^2}
\end{eqnarray}
where
\begin{equation}
A_{s1,t1}=\frac{n_{s,t}\langle \tau _{s,t}\rangle m_{p}}{n_{p}\langle \tau
_{p}\rangle m_{s,t}}.  \label{21}
\end{equation}
and
\begin{equation}
A_{s2,t2}=\frac{g_{s,t}n_{s,t}\langle \tau _{s,t}^{2}\rangle m_{p}}{%
g_{p}n_{p}\langle \tau _{p}^{2}\rangle m_{s,t}}.  \label{22}
\end{equation}

With the use of eqns.(32) and (33) we arrive at the Lorenz number
\begin{eqnarray}
&&L=D_{1}\left[L_{p}+4A_{s1}L_{s}+4A_{t1}L_{t}\right]+ \\
&&D^2_{1}[A_{s1}(2\Gamma _{p}-\Gamma _{s}+\Delta /T)^{2}+  \nonumber \\
&&A_{t1}(2\Gamma _{p}-\Gamma _{t}+(\Delta -J)/T)^{2}+  \nonumber \\
&&4A_{s1}A_{t1}(\Gamma _{t}-\Gamma _{s}+J/T)^{2}]\}  \nonumber
\end{eqnarray}
and the Hall Lorenz number
\begin{eqnarray}
L_{H} &=&D_{2}(L_{p}+4A_{s2}L_{s}+4A_{t2}L_{t})+D_{1}D_{2}^{2}\times  \\
&&\lbrack (4A_{s1}^{2}+2A_{s2})(\Delta /T+2\Gamma _{p}-\Gamma
_{s})^{2}+
\nonumber \\
&&4A_{t1}^{2}((J^{2}+\Delta ^{2})/T^{2}+  \nonumber \\
&&(2\Gamma _{p}-\Gamma _{t})(2(\Delta -J)/T+2\Gamma _{p}-\Gamma _{t}))-
\nonumber \\
&&8A_{t1}A_{s2}(\Delta /T+2\Gamma _{p}-\Gamma
_{s})(J/T+\Gamma_{t}-\Gamma_{s})+  \nonumber \\
&&16(A_{t1}^{2}A_{s2}+A_{s1}^{2}A_{t2})(J/T+\Gamma _{t}-\Gamma
_{s})^{2}+
\nonumber \\
&&2A_{t2}((\Delta/T +2\Gamma _{p}-J/T)^{2}+  \nonumber \\
&&\Gamma _{t}(1+2(J-\Delta )/T-4\Gamma _{p}))+  \nonumber \\
&&8A_{s1}A_{t2}(\Gamma _{s}((\Delta -J)/T-2\Gamma _{t})+  \nonumber \\
&&(J/T+\Gamma _{t})((J-\Delta )/T+\Gamma _{t})-  \nonumber \\
&&2\Gamma _{p}(J/T+\Gamma _{t}-\Gamma _{s}))+
4A_{s1}A_{t1}\times   \nonumber \\
&&(8\Gamma _{p}^{2}+\Gamma _{s}((J-2\Delta )/T+\Gamma
_{t})-2(\Delta/T)\times   \nonumber \\
&&((J+\Delta)/T +\Gamma _{t})-4\Gamma_p(J/T-\Gamma _{s}+\Gamma
_{t}-2\Delta /T))]   \nonumber
\end{eqnarray}
for a strongly-coupled electron-phonon system in the bipolaronic regime,
where we have introduced the dimensionless parameters
\begin{equation}
\Gamma _{\alpha }=\frac{\int_{0}^{\infty }dEE^{2}\tau _{\alpha }(E)\partial
f_{0}^{\alpha }/\partial E}{T\int_{0}^{\infty }dEE\tau _{\alpha }(E)\partial
f_{0}^{\alpha }/\partial E},  \label{10}
\end{equation}
\begin{equation}
\gamma _{\alpha }=\frac{\int_{0}^{\infty }dEE^{3}\tau _{\alpha }(E)\partial
f_{0}^{\alpha }/\partial E}{T^{2}\int_{0}^{\infty }dEE\tau _{\alpha
}(E)\partial f_{0}^{\alpha }/\partial E}.
\end{equation}
and $D_{1,2}=(1+4A_{s1,2}+4A_{t1,2})^{-1}$. Also
\begin{eqnarray}
L_{p} &=&(\gamma _{p}-\Gamma _{p}^{2}), \\
L_{s,t} &=&(\gamma _{s,t}-\Gamma _{s,t}^{2})/4.
\end{eqnarray}

In the limit of a pure polaronic system (i.e. $A_{t}$=$A_{s}=0$) the Lorenz
number and Hall Lorenz number is
\begin{equation}
L=L_{H}=L_{p}.
\end{equation}

If polarons are degenerate then $L_{p}=L_{0}=\pi ^{2}/3$ for any elastic
scattering and any dimensionality. Indeed for a majority of metals $L$ falls
in the region of approximately $(3.1-3.3)$ as expected in a degenerate Fermi
liquid. However $L_{p}$ depends on the dimensionality and the scattering
mechanism for non-degenerate polarons. In the opposite limit we have a pure
singlet or triplet bipolaronic system (i.e. $A_{t}=A_{p}=0$ or $%
A_{s}=A_{p}=0 $). We obtain from eqns.(41) and (42)
\begin{equation}
L=L_{H}=L_{s,t}
\end{equation}
respectively, which is about 6 times smaller than $L_{0}$. The
last equation is expected for a charged Bose gas as noted by Mott
and one of the authors (ASA)\cite{NEV}. In the general case our
final equations, eqn.(41) and (42), yield Lorenz numbers that
differ significantly from both limits at finite temperatures. The
main difference originates from the extra terms, which describe an
interference of polaron and bipolaron contributions to the heat
flow. In the low-temperature regime, $T\ll J$ and $\Delta $, this
contribution is exponentially small because the densities of
triplet pairs and single polarons are small. However, this
contribution becomes important in the intermediate temperature
range $T_{c}<T<T^{\ast }$ because of the combination of the
factors ($J/T)^{2}$ and $(\Delta /T)^{2}$ with the exponential
form of the number densities. The contribution appears as the
result of the recombination of a pair of polarons into triplet and
singlet bound states at the cold end of the sample, which is
reminicent to the contribution of the electron-hole pairs to the
heat flow in semiconductors \cite{ANSE}.
\begin{figure}[h]
\centering \epsfig{file=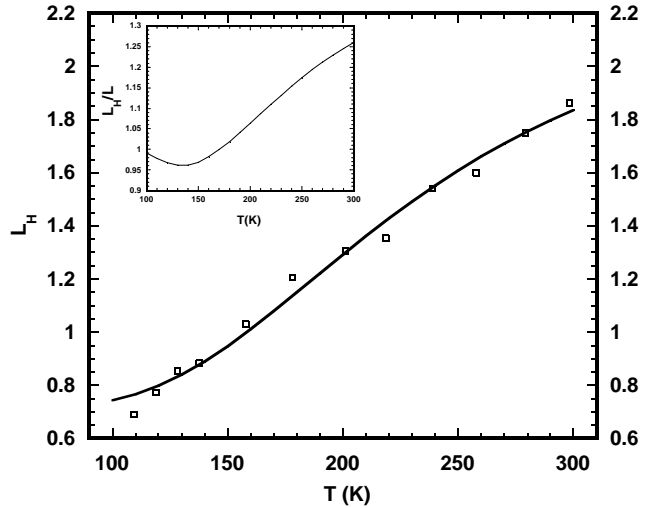,bbllx=85, bblly=186, bburx=523,
bbury=540,width=3.4in} \caption{$L_{H}$ which fits nicely to
experiment ($YBa_{2}Cu_{3}O_{6.95}$) using $\Theta _{b}/\Theta
_{p}=0.36$. This experiment showed a clear violation of the W-F
law. The inset gives the ratio of Hall Lorenz number to Lorenz
number} \label{fig.3}
\end{figure}
Our expressions for the Lorenz numbers eqns.(41,42) anticipate its
values to deviate from $L_{0}/4$ for our system of predominantly
charged Bose particles. Here the $\Gamma $ terms are
representative of a type of scattering mechanism(s), which can be
a combination of types. In the non-degenerate system it is given
by
\begin{equation}
(\gamma -\Gamma ^{2})=\Gamma =(r+2)  \label{33}
\end{equation}
where $r$ is related to the energy dependence of the scattering time
\begin{equation}
\tau \propto E^{r}.  \label{34}
\end{equation}

\section{Numerical analysis and the phonon thermal conductivity}

Here we have shown that the present model fits the Hall Lorenz number, $%
L_{H} $, measured by Zhang et al.\cite{ZHANG}. From these fits the
phonon thermal conductivity can also be extracted.

The charge and
spin gap value was estimated by Mihailovic et al.\cite{mickab} for
optimally doped $YBa_{2}Cu_{3}O_{6+x}$ (x is the number of doped
oxygen ions), giving $\Delta /2=675K$ and $J=150K$, in their
systematic analysis of charge and spin spectroscopies. According
to Ref.\cite{BRAT} the main scattering mechanism above $T_{c}$ is
the particle-particle collisions which gives a relaxation time
$\tau _{s,t,p}\propto 1/T^{2}$. The chemical potential is pinned near the mobility edge, so that
$y=\exp (\mu /T)\approx 0.6$ in a wide temperature range, if the
number of localised states in the random potential is about the
same as the number of bipolarons\cite{BRAT}. In
$YBa_{2}Cu_{3}O_{6+x}$ every excess oxygen ion $x$ can localise a
bipolaron so this approximation is reasonable. As a result, there
is only one fitting parameter in $L_{H}$, Eq.(42) which is the
ratio of the bipolaron and polaron Hall angles $\Theta _{b}/\Theta
_{p}$, where $\Theta _{\alpha }=qB\tau _{\alpha }/m_{\alpha }$ and
we assume $\Theta _{s}=\Theta _{t}=\Theta _{b}$.

The model gives a good fit, as shown in Fig.3, with a reasonable
value of $\Theta _{b}/\Theta _{p}=0.36$. By using the same single
parameter for $L$, as was used to fit $L_{H}$ to the experimental
data, we can see that the ratio of the Lorenz numbers varies with
temperature with $L$ being larger at lower temperatures and
$L_{H}$ growing at higher temperatures (inset in fig. 3).

The model also describes the (quasi) linear in-plane resistivity,
Hall resistivity $T^{2}$-dependence and the Hall ratio, as
observed in the cuprates (fig 4).
\begin{figure}[h]
\centering \epsfig{file=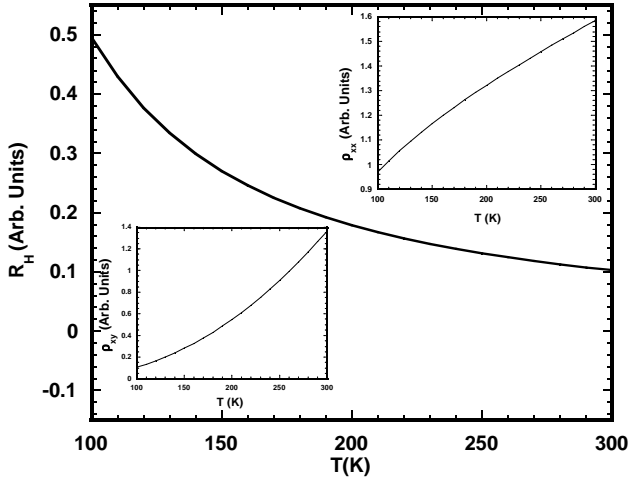,bbllx=71, bblly=210, bburx=520,
bbury=564,width=3.4in} \caption{T dependence of $\rho_{xx}$,
$\rho_{xy}$ and $R_H$ for optimally doped cuprates.} \label{fig.4}
\end{figure}
As mentioned earlier the extraction of the phonon thermal
conductivity from experiment has proven difficult and
inconclusive. Here we can extract this quantity by using our
theoretically calculated $L$ (lower inset of fig. 5) and using
experimental resistivity data \cite{ANDO2} (upper inset of fig.
5). This allows us to find $\kappa_{el}$ in the normal state, fig.
5. Using the same parameters as for the $L_{H}$ fittings, $L$ is
seen to violate the W-F law (lower inset in fig. 5). One can see
that the electronic thermal conductivity is therefore very weakly
T-dependent. These results are similar to the findings by Takenaka
et al and Salamon et al.
\begin{figure}[h]
\centering \epsfig{file=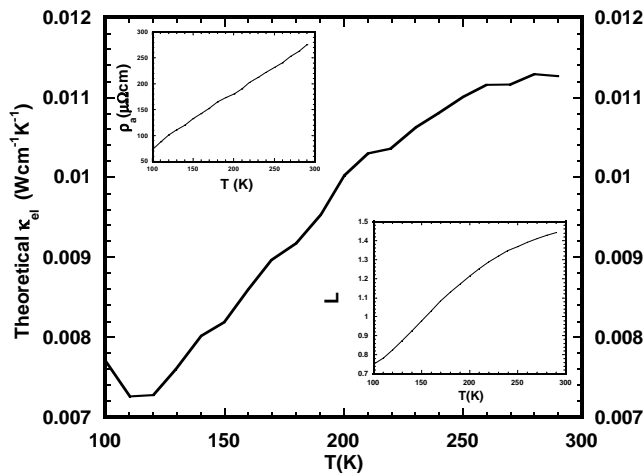,bbllx=69, bblly=194, bburx=535,
bbury=538,width=3.4in} \caption{Normal state
$\kappa_{el}(=LT/\rho$) in optimally doped cuprates calculated
using the theoretical $L$ (lower inset) and experimental
resistivity data (upper inset).} \label{fig.5}
\end{figure}
To find $\kappa_{ph}$ we take for example Minami and Cooper \cite{mina}
thermal conductivity data on $YBa_{2}Cu_{3}O_{6.93}$ and subtract
our $\kappa_{el}$. The results are shown in fig. 6 implying that
the T-dependence of $\kappa$ is predominately due to the phonon
contribution. We believe that this method is the only way to
extract $\kappa_{ph}$ reliably. The lattice contribution to the
diagonal heat flow appears to be much higher than it is
anticipated in the framework of any Fermi-liquid model.

\section{Conclusions}

Recent measurements by Proust et al\cite{PROU} on $Tl_{2}Ba_{2}Cu0_{6+\delta
}$ have suggested that the Wiedemann-Franz law holds perfectly well in the
overdoped region and therefore conclude that the Fermi-liquid prevails at
this doping. Alexandrov and Mott \cite{SANF}, suggested that there might be
a crossover from Bose-Einstein condensation to a BCS-like superconductivity
across the phase diagram. Thus the results of Proust et al are compatible
with the bipolaron picture. If the Fermi liquid does exist at overdoping
then it is likely that the heavy doping causes an "overcrowding effect"
where the polarons find it difficult to form bipolarons due to the larger
number of competing holes\cite{SANF}.
\begin{figure}[h]
\centering \epsfig{file=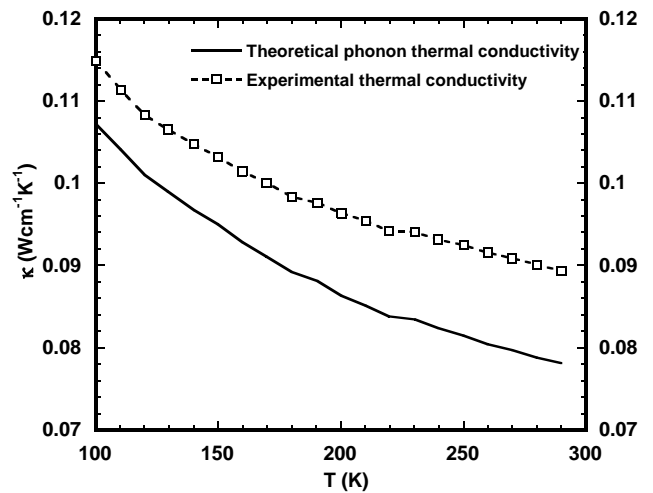,bbllx=71, bblly=210, bburx=524,
bbury=564,width=3.4in} \caption{Deduced $\kappa_{ph}$ and total
$\kappa$ of Minami et al thermal conductivity data on
$YBa_{2}Cu_{3}O_{6.93}$ in the normal state.} \label{fig.6}
\end{figure}
Here we have shown that by the necessary inclusion of thermally
excited polarons and triplets as the temperature rises, the
bipolaronic gas can predict the Hall Lorenz number as found
experimentally. The fits here have been on optimally doped
cuprates, however it is most likely that this model will work in
the underdoped as well. Also this analysis, along with
experimental data, allowed us to estimate the phonon contribution
to the thermal conductivity which has remained elusive in recent
years. The interference of the polaron and bipolaron contributions
to the energy flow breaks down the Wiedemann-Franz law and results
in the unusual temperature dependence of the Lorenz numbers. This
work further validates the Bipolaronic model and it is our belief
that the superconducting state heat transport can be described by
it as well.

This work was supported by the Leverhulme Trust (grant F/00261/H)
and by the EPSRC UK (grant R46977). We would like to thank Y.
Zhang and P.W. Anderson for helpful discussions of some
experimental and theoretical aspects.


\begin{references}
\bibitem{CHAK}  S. Chakravarty, R.B. Laughlin, D.K. Morr and C. Nayak, Phys.
Rev. B, {\bf 63}, 094503 (2001).

\bibitem{SANF}  A.S. Alexandrov and N.F. Mott, {\it High Temperature
Superconductors and Other Superfluids}, (Taylor and Francis, London, 1994).

\bibitem{KIVE}  V.J. Emery and S.A. Kivelson, Nature, {\bf 374}, 434 (1995).

\bibitem{PWA}  P.W. Anderson, {\it The Theory of Superconductivity in the
High T$_{c}$ Cuprates}, (Princeton University Press, Princeton, 1997).

\bibitem{ZHAN}  S.C. Zhang, Science, {\bf 275}, 4126 (1997).

\bibitem{PINE}  D. Pines, A.V. Balatsky and P. Monthoux, Phys. Rev. Lett,
{\bf 67}, 3448 (1991).

\bibitem{ZHAO}  G. Zhao, M.B. Hunt, H. Keller and K.A. Muller, Nature, {\bf %
385}, 236 (1997).

\bibitem{mic}  D. Mihailovic, C.M. Foster, K. Voss and A.J. Heeger, Phys.
Rev. B{\bf 42}, 7989 (1990).

\bibitem{ita}  P. Calvani, M.Capizzi, S. Lupi, P. Maselli, A. Paolone, P.
Roy, S-W Cheong, W. Sadowski and E. Walker, Solid State Commun. {\bf 91},
113 (1994).

\bibitem{TIM}  T. Timusk, C.C. Homes and W. Reichardt, in {\it \ Anharmonic
properties of High T$_{c}$ Cuprates}, eds. D.Mihailovic, G. Ruani, E. Kaldis
and K.A. Muller, 171(World Scientific, Singapore, 1995).

\bibitem{COHN}  J.L. Cohn, S. Wolf and T.A. Vanderah, Phys. Rev. B , {\bf 45}%
, 511 (1992).

\bibitem{ega}  T. Egami, J. Low Temp. Phys. {\bf 105}, 791(1996).

\bibitem{LANZ}  A. Lanzara, P.V. Bogdanov, X.J. Zhou, S.A. Kellar, D.L.
Feng, E.D. Lu, T. Yoshida, H. Eisaki, A. Fujimori,K. Kishio, J.I. Shimoyana,
T. Noda, S. Uchida, Z. Hussain and Z.X. Shen, Nature, {\bf 412}, 510 (2001).

\bibitem{CHAI}  A. Chainani, T. Yokoya, T. Kiss, S. Shin, T. Nishio and H.
Uwe, Phys. Rev. B, {\bf 64}, 180509 (2001).

\bibitem{ELIA}  G.M. Eliashberg, Sov.Phys. JETP, {\bf 11}, 696 (1960).

\bibitem{ALEX}  A.S. Alexandrov, Phys. Rev. B, {\bf 46}, 2838 (1992).

\bibitem{LANG}  I.G. Lang and Y.A. Firsov, Sov. Phys. JETP, {\bf 16}, 1301
(1963).

\bibitem{SASH}  A.S. Alexandrov, Russ. J. Phys. Chem., {\bf 57}, 273 (1983).

\bibitem{SAS}  A.S. Alexandrov and E.A. Mazur, Zh. Eksp. Teor. Fiz., {\bf 96}%
, 1773 (1989).

\bibitem{MULL}  J.G. Bednorz and K.A. Muller, Angew. Chem., Int. Ed. Engl.
{\bf 57}, 735 (1988).

\bibitem{ALEXAND}  A.S. Alexandrov, Phys. Rev. B, {\bf 53}, 2863 (1996).

\bibitem{CATL}  C.R.A. Catlow, M.S. Islam and X. Zhang, J. Phys.:Condensed Matter, {\bf 10} No. 3, L49
(1998).

\bibitem{tru}  J. Bonca J and S.A. Trugman, Phys. Rev. B {\bf 64}, 094507
(2001).

\bibitem{alekor}  A.S. Alexandrov and P.E. Kornilovitch, J. Phys.:Condensed
Matter, {\bf 14} No. 21, 5337 (2002).

\bibitem{MOTT}  A.S. Alexandrov and N.F. Mott, J. Supercond (US), {\bf 7},
599 (1994).

\bibitem{ROSS}  J. Rossat-Mignod, L.P. Regnault, P. Bourges, C. Vettier, P.
Burlet and J.Y. Henry, Physica Scripta, {\bf 45}, 74 {1992). }

\bibitem{MOOK}  H.A. Mook and M. Yethiraj, Phys. Rev. Lett, {\bf 70}, 3490
(1993).

\bibitem{mickab}  D. Mihailovic, V.V. Kabanov, K. Zagar, and J. Demsar,
Phys. Rev. B{\bf 60}, 6995 (1999) and references therein.

\bibitem{DEUT}  G. Deutscher, Nature, {\bf 397}, 410 (1999).

\bibitem{AND}  A.S. Alexandrov and A.F. Andreev, EuroPhys. Lett., {\bf 54},
373 (2001).

\bibitem{BRAT}  A.S. Alexandrov, A.M. Bratkovsky and N.F. Mott, Phys. Rev.
Lett, {\bf 72}, 1734 (1994).

\bibitem{SALJ}  A.S.Alexandrov, A.M. Bratkovsky, N.F. Mott and E.K.H Salje,
Physica C, {\bf 215}, 359 (1993).

\bibitem{MULLE}  K.A. Muller, G.M. Zhao, K. Conder and H. Keller, J. Phys.
Cond. Mat., {\bf 10}, L291 (1998).

\bibitem{NDROV}  A.S. Alexandrov, Physica C, {\bf 305}, 46 (1998).

\bibitem{ALEXA}  A.S. Alexandrov, Phys. Rev. B, {\bf 46}, 14932 (1992).

\bibitem{NEV}  A.S. Alexandrov and N.F. Mott, Phys. Rev. Lett, {\bf 71},
1075 (1993).

\bibitem{SASHA}  A.S. Alexandrov, Doctoral thesis, Moscow Engineering
Physics institute (1984).

\bibitem{DENT}  A.S.Alexandrov and C.J. Dent, Phys. Rev. B, {\bf 60}, 15414
(1999).

\bibitem{JUN}  A. Junod, {\it Studies of High-Temperature Superconductors
Vol. 19}, eds. A. Narlikar, 1 (Nova science, Commack, NY, 1996).

\bibitem{DROV}  A.S. Alexandrov, W.H. Beere, V.V. Kabanov and W.Y Liang,
Phys. Rev. Lett., {\bf 79}, 1551 (1997).

\bibitem{JUNO}  A. Junod, A. Erb and C Renner, Physica C, {\bf 317-318}, 333
(1999).

\bibitem{ZHANG}  Y. Zhang, N.P. Ong, Z.A. Xu, K. Krishana, R. Gagnon, and L.
Taillefer, Phys. Rev. Lett., {\bf 84}, 2219 (2000).

\bibitem{kkl}  K.K. Lee, A.S. Alexandrov and W.Y Liang, Cond-mat/0211665
(2002).

\bibitem{TAKE}  K. Takenaka, Y. Fukuzumi, K. Mizuhashi, S. Uchida, H. Asaoka
and H. Takei, Phys. Rev. B, {\bf 56}, 5654 (1997).

\bibitem{SALA}  R.C. Yu, M.B. Salamon, J.P. Lu and W.C. Lee, Phys. Rev.
Lett., {\bf 69}, 1431 (1992).

\bibitem{HILL}  R.W. Hill, C. Proust, L. Taillefer, P. Fournier and R.L.
Greene, Nature, {\bf 414}, 711 (2001).

\bibitem{ANDO}  J. Takeya, Y. Ando, S. Komiya and X.F. Sun, Cond-mat/0108055
(2002).

\bibitem{kiv}  E. W. Carlson, V. J. Emery, S. A. Kivelson, and D. Orgad,
cond-mat/0206217 and references therein.

\bibitem{pop}  A. J. Leggett, Physica Fennica {\bf 8,} 125 (1973); J Stat.
Phys. {\bf 93}, 927 (1998);

V. N. Popov, {\it Functional Integrals and Collective Excitations}
(Cambridge: Cambridge University Press) (1987).

\bibitem{aleF}  A.S. Alexandrov, Physica C {\bf 363}, 231 (2001).

\bibitem{mul2}  K.A. Muller, Phil. Mag. Lett, {\bf 82}, 279 (2002).

\bibitem{ZHAO1}  G.M. Zhao, Phys. Rev. B, {\bf 64}, 024503 (2001).

\bibitem{ANSE}  A. Anselm, {\it Introduction of Semiconductor Theory} ,
(Prentice and Hall, New Jersey, 1981).

\bibitem{ANDO2}  Kouji Segawa and Yoichi Ando, Phys. Rev. Lett., {\bf 86},
4907 (2001).

\bibitem{mina}  H. Minami and J.R. Cooper, Private communication.

\bibitem{PROU}  C. Proust, E. Boakin, R.W. Hill, L. Taillefer and A.P.
Mackenzie, Cond-Mat/0202101 (2002).
\end{references}
\end{document}